\documentclass[preprint,numbers,11pt]{elsarticle}
\usepackage{amssymb, amsmath, fullpage, setspace}
\usepackage[pdftex,colorlinks]{hyperref}

\newcommand{\EQ}[1]{\begin{align}\begin{split} #1 
\end{split}\end{align}}
\newcommand{\eq}[1]{\begin{equation}{ #1 
}\end{equation}}

\journal{}

\begin{document}

\begin{frontmatter}

\title{Cluster Alphabets from Generalized Worldsheets: A Geometric Approach to Finite Types}

\author[a]{Peng Zhao}
\author[b]{and Yihong Wang}

\affiliation[a]{
organization={Joint School of the National University of Singapore and Tianjin University, International Campus of Tianjin University}, 
city={Fuzhou}, 
postcode={350207}, 
country={China}}
\affiliation[b]{
organization={
Laboratoire d’Annecy-le-Vieux de Physique Theorique (LAPTh), CNRS and Universite Savoie Mont-Blanc},
city={Annecy},
postcode={74940},
country={France}
}

\begin{abstract}
We provide a systematic derivation of cluster alphabets of finite types. The construction is based on a geometric realization of the generalized worldsheets by gluing and folding a pair of polygons. The cross ratios of the worldsheet $z$ variables are evolved using the $Y$-system equations. By a new gauge choice, we obtain a simpler set of cluster alphabets than the known ones. 
\end{abstract}

\end{frontmatter}

\section{Introduction and Summary}

The search for a geometric description and a simple set of variables has guided the study of scattering amplitudes in quantum field theory and string theory. Historically, the Veneziano amplitude consistent with the Regge poles and crossing symmetry was written down first, then extended to $n$-point amplitudes, and the notion of a worldsheet swept out by the motion of strings emerged only later.

The generalization of the Veneziano amplitude to $n$ mesons is expressed as  \cite{Chan:1969xg, Chan:1969ex, Koba:1969rw}
\eq{
I_{n} = \left(\prod^{n-3}_{(i,j)}\int_0^1 d\log \frac{u_{i,j}}{1-u_{i,j}} \right) \prod^{n}_{i,j} u_{i,j}^{\alpha' X_{i,j}}\,.
\label{stringyintegral}
} 
Here the integral is over the $n-3$ compatible resonances, $\alpha'$ is the Regge slope, and $X_{i,j}$ are functions of the Mandelstam variables $s_{i,j}$. It was soon realized that the $u$ variables may be written as cross ratios of the Koba-Nielsen $z$ variables \cite{Koba:1969kh}.
\eq{
u_{i,j} = \frac{z_{i-1}-z_{j}}{z_{i-1}-z_{j-1}} \frac{z_{i}-z_{j-1}}{z_{i}-z_{j}}\label{crossratio}
\,.}
This leads to the expression of the integration measure in the Parke-Taylor form $\prod_{i=1}^n dz_i/{(z_{i}-z_{i+1})}$ and the Koba-Nielsen factor $|z_i - z_j|^{\alpha's_{i,j}}$ familiar in modern textbooks, with the residual gauge symmetry used to fix the positions of three points, e.g., $z_1 \to -1, z_2 \to 0, z_n \to \infty$. It is now recognized that this integral describes the tree-level amplitude of open strings, whose worldsheet is a disc with $n$ marked points at the boundary. The string amplitude enjoys properties such as crossing symmetry, factorization, and Regge behavior.

More recently, the factorization property of the string integral was put on the center stage to define a class of generalized string integrals associated with Dynkin diagrams \cite{Arkani-Hamed:2019mrd}. The so-called cluster string integrals factorize at the poles that correspond to the boundaries of the configuration space of $u$ variables \cite{Arkani-Hamed:2019plo, zbMATH07431231}. For example, the $A_{n-3}$ integral \eqref{stringyintegral} factorizes into an $A_{n-k-2}$ integral and an $A_{k-2}$ integral at its poles. The factorization property is reflected in the geometry of the generalized associahedra \cite{zbMATH02068688} and the integrals are interpreted as volume forms. However, like the multimeson amplitudes, the integrals are written in terms of the $u$ variables as a generalization of \eqref{stringyintegral}. It was not clear what the underlying worldsheet picture is.

A second motivation for this work comes from the structure of field-theory amplitudes. The amplitudes are expressed in terms of generalized polylogarithms. The cluster bootstrap program attempts to constrain the form of the amplitude using a set of symbol alphabets \cite{Dixon:2011pw, Golden:2013xva, Drummond:2014ffa}. In a related development \cite{Chicherin:2020umh}, a class of alphabets based on cluster algebras of finite type was proposed using birational maps from the kinematic variables:

\EQ{
\Phi_{A_{n-3}} &= \bigcup_{3 \le i \le n-1} \{z_i,\, 1+z_i \} \cup \bigcup_{3 \le i < j \le n-1} \{z_{i} - z_{j}\} \,, \\
\Phi_{C_{n-1}} &=\Phi_{A_{n-1}}(z_3, \ldots, z_{n}, z_{n+2})\cup \bigcup_{3\le i \le j \le n} \{ z_i z_j + z_{n+2} \}
\,,\\
\Phi_{D_n} &= \Phi_{A_{n-1}}(z_3, \ldots, z_{n}, z_{n+2}) \cup \{z_{n+3}, 1+z_{n+3}\} \cup \bigcup_{3 \le i \le n} \{z_{i}-z_{n+3}, z_i  + z_{n+2}z_{n+3} \} \\
&\cup \bigcup_{3 \le i < j \le n} \{-z_i+z_j+z_i z_j-z_i z_{n+2}-z_i z_{n+3}+z_{n+2} z_{n+3}\}.
\label{ACDalphabet}
}
It was shown that the Feynman integral for the one-loop Bhabha scattering correspond to the $A_3$ cluster alphabet, a certain six-dimensional hexagon integral to the $D_4$ cluster alphabet, etc. However, the cluster alphabets were found by a clever choice of birational maps, and it was not clear how to derive them for other finite-type cluster algebras. The $A$-type alphabet is the set of gauge-fixed factors in the $u$ variables \eqref{crossratio}, as string amplitudes reduce to field-theory amplitudes in the $\alpha' \to 0$ limit. It was also not clear whether the other alphabets have any geometric origin or if there is an underlying worldsheet at all.

In \cite{He:2021zuv}, a systematic derivation of such variables was proposed based on $Y$ systems, and the results for $D$ types were presented in detail. The strategy is to construct the generalized worldsheets through a ``gluing" construction. We solve the $u$ variables in terms of the worldsheet coordinates. Like in the $A$ type, the elements of an alphabet, called letters, are the factors that appear in the $u$ variables. As it stands, there are more ungauged letters than the number of cluster variables. Upon a choice of gauge, the alphabets are then read off from the factors. The number of letters in a cluster alphabet is shown in Table 1.

\begin{table}[htbp]
\center
\begin{tabular}{|c|c|c|c|c|c|c|c|}
\hline
$A_{n-3}$ & $B_n$/$C_n$ & $D_n$ & $E_6$ & $E_7$ & $E_8$ & $F_4$ & $G_2$ \\
\hline
$n(n-3)/2$ & $n(n+1)$ & $n^2$ & 42 & 70 & 128 & 28 & 8 \\
\hline
\end{tabular}
\caption{The dimensions of finite-type cluster algebras, which equal the number of letters in a cluster alphabet.}
\end{table}

This paper aims to derive the cluster alphabets for all the finite types. Our main results are as follows:
\begin{itemize}
\item Systematic construction of the generalized worldsheet for all finite types.

We provide a systematic derivation of the gluing construction of the exceptional types. We begin by reviewing the gluing construction of the $D_n$ worldsheet. We show how the construction extends to the exceptional types, and derive an explicit cross-ratio representation of all the $E_6$  $u$ coordinates. For the nonsimply laced types, we present the folding map that identifies the worldsheet coordinates. Our results may also be seen as an explicit verification of Zamolodchikov's periodicity conjecture for $Y$ systems \cite{Zamolodchikov:1991et}. 

\item New cluster alphabets. 

In the standard gauge choice, one may obtain the cluster alphabet for $BCD, E_6, F_4, G_2$ types as polynomials of degrees at most 2, 4, 5, 4, respectively. By choosing a different gauge, we produce a new, simpler set of alphabets. We obtain a linear alphabet for $B$ type, quadratic alphabets for $CD$ types, and for $E_6, E_7, E_8, F_4, G_2$ types, polynomials of degrees at most 4, 5, 7, 4, 2, respectively. 

\end{itemize}

\section{The gluing construction}

\subsection{Review on the gluing construction of the $D_n$ worldsheet}

\begin{figure}
\centering
\includegraphics[scale=0.2]{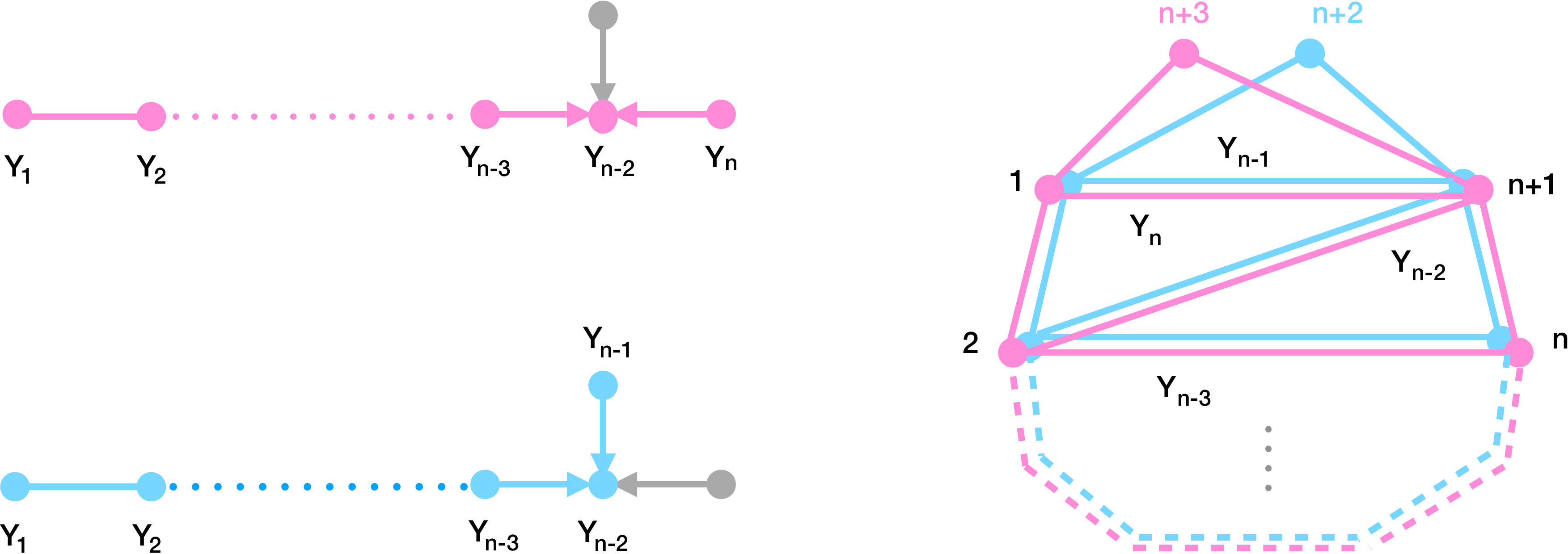}
\caption{
The glued-polygon representation of the $D_n$ worldsheet.}
\label{fig:Dnpolygon}
\end{figure} 

We review the construction of the $D$-type worldsheet based on gluing a pair of $A$-type worldsheets. The basic observation is that the $D_n$ Dynkin diagram can be written as a union of a pair of $A_{n-1}$ Dynkin diagrams, as shown in Fig. \ref{fig:Dnpolygon}. We prepare two $(n+2)$-gons. We will call the first polygon the first sheet and the second polygon the second sheet. The vertices of the polygons can be given any labels but for convenience, we will choose them to be $(1, 2,\ldots, n+1, n+2)$ and $(1,2, \ldots, n+1, n+3)$, respectively. We glue $n+1$ of the common vertices together, leaving the last vertex on each polygon alone. The positions of the vertices $z_1, \ldots, z_{n+3}$ will be our worldsheet variables. We may choose a snake triangulation. Assigning a node to each diagonal and an arrow between two consecutive diagonals ordered counterclockwise around a common vertex, we see that the underlying graph precisely corresponds to a Dynkin diagram of type $D_n$ \cite{Gekhtman_2005}.

Recall that in an $A_n$ worldsheet, the $u$ variables are cross ratios of their respective $z$ variables \eqref{crossratio}. By a reparametrization $u = Y/(1+Y)$, it is found that the cross ratios on the worldsheet satisfy a celebrated set of equations, known as $Y$ systems:
\eq{Y_{i,j} Y_{i+1,j+1} = (1+Y_{i,j+1})  (1+Y_{i+1,j})\,.}
This can be seen as a time-evolution equation that maps $Y_{i,j} \to Y_{i+1,j+1}$ in each time step.

The $Y$-system equations admit a generalization to all finite types \cite{Zamolodchikov:1991et}:
\eq{Y_i(t-1) Y_i(t) = \prod_{j \to i} (1+Y_j(t))^{-C_{i,j}} \prod_{i \to j} (1+Y_j(t-1))^{-C_{i,j}}\,.
\label{Ysystems}
}
Here $C_{i,j}$ is the Cartan matrix of the root system. One may assign an orientation to the edges in the Dynkin diagram such that each node is either a source or a sink. The Zamolodchikov periodicity conjecture states that the solutions to \eqref{Ysystems} are periodic. A solution of $Y$ systems in terms of cross ratios was used to prove the periodicity conjecture for the $A$ type \cite{Volkov_2007}. 

Returning to the worldsheet picture, the $n-2$ diagonals in the triangulation are identified with the initial $Y$ variables for each common node in the Dynkin diagram $Y_{i}(0)$ for $i=1,2, \ldots, n-2$. The diagonals connecting $z_{1}$ with $z_{n+1}$ in the $(1, 2, n+1, n+2)$ and $(1,2,n+1,n+3)$ quadrilaterals provide $Y_{n-1}(0)$ and $Y_{n}(0)$ for the two branched nodes in the Dynkin diagram, respectively. Now we may assign cross ratios to the diagonals in the initial triangulation. The $Y_{i}(t)$ variables at later times are generated according to the $Y$-system equations. This process terminates when the $Y$ variables return to their initial values as guaranteed by periodicity. 

Because the $Y$-system equations are always birational transformations on the $Y$ variables, the new $Y$ variables will always be a rational function of the $z$ variables. Let $z_{i,j} := z_j - z_i$. Remarkably, once we introduce the cubic polynomials
\eq{w_{i,j} =  z_{1,n+3} z_{i,j} z_{n+1,n+2} - z_{1,n+1} z_{i,n+3} z_{j,n+2}\,,}
the corresponding $u$ variables can be written as generalized cross ratios of the $z,w$ factors:
\EQ{
u_{i,\, j} &=\frac{z_{i,j-1}\, z_{i-1,j}}{z_{i,j}\, z_{i-1,j-1} }\,, \quad u_{j,\, i} = \frac{w_{i,j-1}\, w_{i-1,j}}{w_{i,j}\, w_{i-1,j-1}} \,, \\
u_{i} &=\frac{z_{i,n+3} \, w_{i-1,i} }{z_{i-1,n+3} \, w_{i,i}}\,, \qquad \widetilde u_{i} =\frac{z_{i,n+2}\,w_{i-1,i}}{z_{i-1,n+2}\,w_{i,i}}\,,
}
for $n+1 \ge  i > j > 1$. Here $u_i$ and $\widetilde u_i$ correspond to the two branched nodes in the Dynkin diagram. Because the $D_n$ worldsheet is constructed from gluing a pair of $A_{n-1}$ worldsheets, one may think of $u_{i,j}$ with $i>j$ as the cross-ratio coordinates of $z$'s of the first sheet, and $u_{i,j}$ with $i<j$ as the cross-ratio coordinates of $w$'s of the second sheet.

We shall denote the collection of polynomial factors that appear in the $u$ variables as an ``ungauged alphabet." The ungauged $D_{n}$ alphabet is
\eq{\bigcup_{1\le i \le n+1}\{z_{i,n+3}\}\,\cup \bigcup_{1\le i < j \le n+2} \{z_{i,j}\}\,\cup \bigcup_{2 \le i < j \le n}\{w_{i,j} \}\,. }
There are $n^2 + n + 3$ independent variables. Upon gauge fixing $z_1 \to -1, z_2 \to 0, z_{n+1} \to \infty$, $n+3$ variables corresponding to $z_{1,2}$ and $z_{i,n+1}$ for all $i \ne n+1$ are removed, and we obtain the $n^2$ letters \eqref{ACDalphabet}.

While the $u$ variables are written nicely as generalized cross ratios, the interpretation of the $w$ variables remains mysterious. Here we provide a new, determinant representation:
\eq{
w_{i,j} = 
\det\left(
\begin{array}{ccc}
 1 & 1 & 1 \\
z_i+z_{n+1} & z_1 + z_j  & z_{n+2}+z_{n+3} \\
z_i z_{n+1} & z_1 z_j  & z_{n+2} z_{n+3} \\
\end{array}
\right)
\,.
\label{det}}
It is symmetric on the pairs of indices $(i,n+1)$, $(1,j)$, $(n+2,n+3)$, but is antisymmetric when the pairs are exchanged, much like the symmetries of a Riemann tensor. 

\subsection{The construction of $E_n$ worldsheets}
\begin{figure}
\center
\includegraphics[scale=0.2]{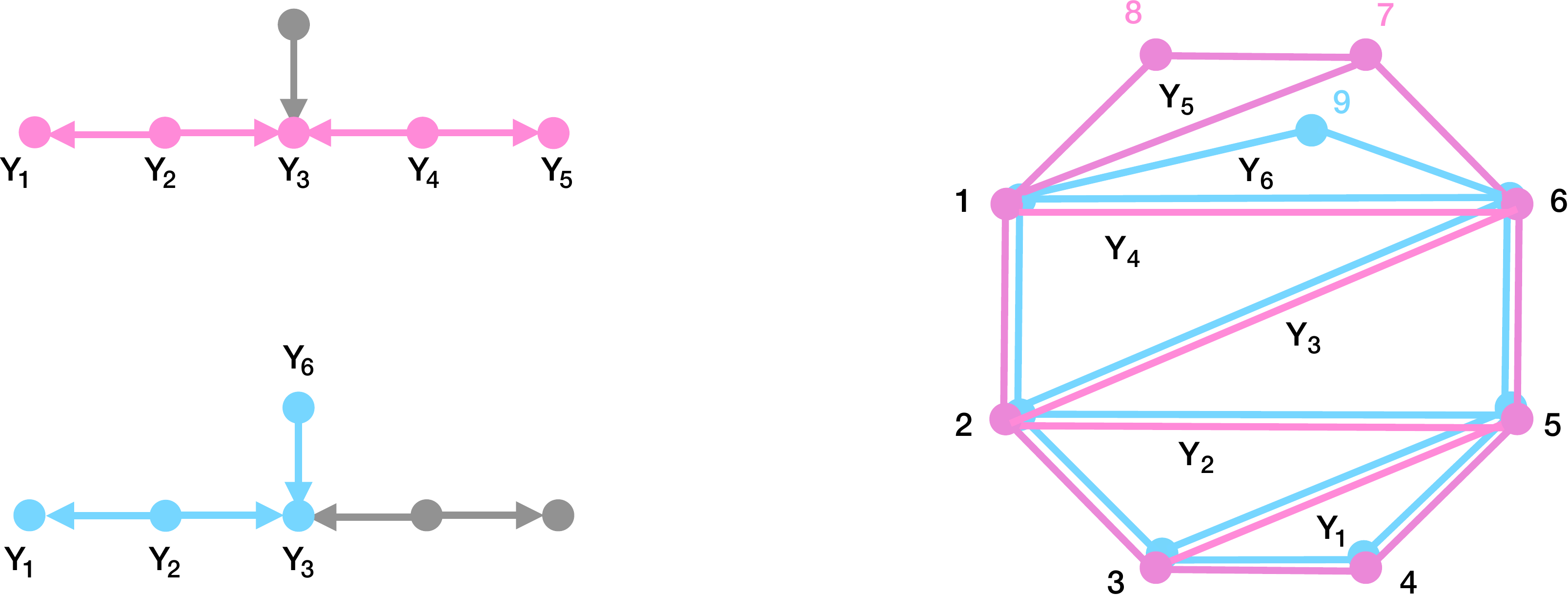}
\caption{The glued-polygon representation of the $E_6$ worldsheet. }
\label{fig:E6polygon}
\end{figure}

Consider an $E_n$-type Dynkin diagram, where $n=6,7,8$. It may be written as a union of $A_{n-1}$ and $A_{n-2}$ diagrams. We prepare a $(n+2)$-gon and a $(n+1)$-gon and glue $n$ of the common vertices, leaving one vertex on the first polygon and two vertices on the second polygon free. We work out the $E_6$ example explicitly, as shown in Fig. \ref{fig:E6polygon}. The initial set of variables are 
\EQ{
Y_{i}(0) 
&=\left\{Y_{3,5}, Y_{2,5}, Y_{2,6}. Y_{6,1}, Y_{7,1}, {\widetilde Y}_{6,1}\right\}
\\
&=\left\{\frac{z_{2,5} z_{3,4}}{z_{2,3} z_{4,5}},\frac{z_{1,5} z_{2,4}}{z_{1,2} z_{4,5}},\frac{z_{1,6} z_{2,5}}{z_{1,2} z_{5,6}},\frac{z_{1,5} z_{6,8}}{z_{1,8} z_{5,6}},\frac{z_{1,6} z_{7,8}}{z_{1,8} z_{6,7}},\frac{z_{1,5} z_{6,9}}{z_{1,9} z_{5,6}}\right\}
\,.
}
The $Y$ variables are written in terms of the $z$ variables as 
\eq{Y_{i,j} = \frac{z_{i-1,j}\, z_{i,j-1}}{z_{i-1,i}\, z_{j-1,j}} \,.}
Note that on the first sheet, the vertex that comes before $z_1$ is $z_8$; on the second sheet, the vertex that comes before $z_1$ is $z_9$.

We evolve the $Y$-system equations \eqref{Ysystems} as before, generating all the $Y_{i}(t)$ at later times. Among the nonlinear factors appearing in this parametrization of the $E_{6}$ $Y$-system, there are $12$ cubic polynomials of the form
\EQ{
w_{i,j}^{E} &= z_{1,n+1} z_{i,j}  z_{n,n+3} - z_{1,n} z_{i,n+1} z_{j,n+3}\,, \\
\widetilde w_{i,j}^{E} &= z_{1,n+2} z_{i,j}  z_{n,n+3} - z_{1,n} z_{i,n+2} z_{j,n+3} \,.\\
}
Note that $w_{i,j}^{E}$ for $E_6$ is slightly different from $w_{i,j}$ for $D_n$ due to a difference in the labels. 
There are also four quartic polynomials of the form
\eq{
w_{i,j,k} = z_{1,i} z_{j,n+2} z_{k,n+1}  z_{n,n+3} - z_{1,n+3} z_{n+2,i} z_{n+1,j} z_{n,k}\,,
}
and a sextic polynomial
\EQ{
w_{i,j,k,l} &= z_{1,n+3} z_{1,n} z_{i,n+2} z_{j,n+1} z_{k,n+3} z_{l,n}-z_{1,n+3} z_{1,n} z_{n,n+3} z_{i,n+2} z_{j,k} z_{l,n+1} \\
&+z_{1,i} z_{1,n} z_{n,n+3} z_{j,n+3} z_{k,n+2} z_{l,n+1}+z_{1,n+2} z_{1,n+3} z_{n,n+1} z_{n,n+3} z_{i,l} z_{j,k}\,. 
}
The indices are taken to lie in $2 \le i<j<k<l \le 5$ so the last polynomial is simply $w_{2,3,4,5}$. When some of the indices are allowed to coincide, the sextic polynomial factorizes into a product of the lower-order polynomials, e.g.,
\eq{
w_{i,i,j,k} = w_{i,k}^{E} \widetilde w_{i,j}^{E} , \qquad w_{i,j,j,k} = z_{1,n}z_{j,n+3} w_{i,j,k}\,.
}
This allows us to write the $Y$ variables, or equivalently the $u$ variables, of $E_6$ compactly as generalized cross ratios\footnote{The $u$ variables for $E_6$ can alternatively be realized by the Grassmannian cluster algebra G(4,7) \cite{Drummond:2018dfd}.}:
\EQ{
&u_{i}(t) = \\
&\begin{pmatrix}
\frac{z_{2,5} z_{3,4}}{z_{2,4} z_{3,5}} & \frac{z_{3,6} z_{4,5}}{z_{3,5} z_{4,6}} & \frac{w_{2,2,2,4} z_{5,6}}{w_{2,2,2,5} z_{4,6}} & \frac{w_{2,3,3,3} w_{2,3,3,5}}{w_{2,2,3,5} w_{3,3,3,3}} & \frac{z_{4,8} w_{3,4,4,4}}{z_{3,8}w_{4,4,4,4} } & \frac{z_{4,7} w_{4,4,4,5} }{z_{5,7}w_{4,4,4,4} } & \frac{z_{5,8} z_{6,7}}{z_{5,7} z_{6,8}}
\\
\frac{z_{1,5} z_{2,4}}{z_{1,4} z_{2,5}} & \frac{z_{2,6} z_{3,5}}{z_{2,5} z_{3,6}} & \frac{z_{4,6}w_{2,2,2,3} }{z_{3,6}w_{2,2,2,4} } & \frac{w_{2,2,2,5} w_{2,3,3,4}}{w_{2,2,2,4} w_{2,3,3,5}} &  \frac{w_{2,3,4,5} w_{3,3,3,5} }{w_{2,3,3,5}w_{3,3,4,5}} &  \frac{w _{3,4,4,5} w _{4,4,4,4}}{w _{3,4,4,4} w _{4,4,4,5}}  & \frac{w_{3,5,5,5} w_{4,5,5,5}}{w_{3,4,5,5} w_{5,5,5,5}} \\
 \frac{z _{1,6} z _{2,5}}{z _{1,5} z _{2,6}} & \frac{z _{3,6} w _{2,2,2,2}}{z _{2,6} w _{2,2,2,3}} & \frac{w _{2,2,2,4} w _{2,3,3,3}}{w _{2,2,2,3} w _{2,3,3,4}} &  \frac{w_{2,3,3,5} w_{2,3,4,4}}{w_{2,3,3,4} w_{2,3,4,5}} & \frac{w_{2,4,4,5} w_{3,3,4,5}}{w_{2,3,4,5} w_{3,4,4,5}} & \frac{w _{3,5,5,5} w _{4,4,4,5}}{w _{3,4,4,5} w _{4,5,5,5}} & \frac{z _{1,4} w _{5,5,5,5}}{z _{1,5} w _{4,5,5,5}} \\ 
\frac{z _{1,5} z _{6,8}}{z _{1,6} z _{5,8}} & \frac{z _{1,7} z _{2,6}}{z _{1,6} z _{2,7}} &  \frac{w_{2,2,2,3}w_{2,2,2,4} }{w_{2,2,2,2} w_{2,2,3,4}} & \frac{w_{2,3,3,4} w_{3,3,3,3}}{w_{2,3,3,3} w_{3,3,3,4}} &  \frac{w_{2,3,4,5} w_{3,4,4,4}}{w_{2,4,4,5} w_{3,3,4,4}} & \frac{w _{2,5,5,5} w _{3,4,4,5}}{w _{2,4,4,5} w _{3,5,5,5}} & \frac{z _{1,3} w _{4,5,5,5}}{z _{1,4} w _{3,5,5,5}} \\ \frac{z _{1,6} z _{7,8}}{z _{1,7} z _{6,8}} & \frac{z _{1,8} z _{2,7}}{z _{1,7} z _{2,8}} & \frac{z _{3,8} w _{2,3,3,3}}{z _{2,8} w _{3,3,3,3}} & \frac{ z_{3,7}w_{3,3,3,4}}{z_{4,7}w_{3,3,3,3} } & \frac{w_{2,4,4,5} w_{4,4,4,5}}{w_{2,4,5,5} w_{4,4,4,4}} & \frac{z _{1,2} w _{3,5,5,5}}{z _{1,3} w _{2,5,5,5}} & \frac{z _{1,4} z _{2,3}}{z _{1,3} z _{2,4}} \\
\frac{z _{1,5} z _{6,9}}{z _{1,6} z _{5,9}}  & \frac{z _{1,9} z _{2,6}}{z _{1,6} z _{2,9}} & \frac{z _{2,7} w _{2,2,2,3}}{z _{3,7} w _{2,2,2,2}} & \frac{w_{2,3,3,4}w_{3,3,3,4} }{w_{2,3,4,4}w_{3,3,3,3} } & \frac{w_{2,3,4,5} w_{3,4,4,4}}{w_{2,4,4,4} w_{3,3,4,5}} & \frac{w_{3,4,4,4} w_{3,4,4,5}}{w_{3,3,4,5}w_{4,4,4,4} } &  \frac{z _{5,8} w _{4,5,5,5}}{z _{4,8} w _{5,5,5,5}} 
\end{pmatrix}\,.
}
Unlike the $D_n$ case, the cross ratios involving $w_{i,j,k,l}$ are not unique and can be transformed using the identities
\eq{
\frac{w_{i,i,j+1,k}\, w_{i,i,j,k+1}}{w_{i,i,j,k}\, w_{i,i,j+1,k+1}} = \frac{w_{i+1,j,k,k}\, w_{i,j+1,k,k}}{w_{i,j,k,k}\, w_{i+1,j+1,k,k}} = 1 \,.
}

In the standard gauge choice $(z_1 = -1, z_2 = 0, z_{6} = \infty)$, the $E_6$ alphabet consists of 42 letters that are polynomials with a degree of at most 4:
\EQ{
&\Phi_{E_6} = \Phi_{A_5} (z_3, z_4, z_5, z_7, z_8) \cup \{z_9, 1+z_9\} \cup \bigcup_{3\le i \le 5} \{z_{i,9}, \, z_{i}+z_7 z_9, \, z_{i}+z_8 z_9 \} \, \cup   \\
&\bigcup_{3\le i < j \le 5} \{-z_i+z_j+z_i z_j- z_i z_7- z_i z_9 + z_7 z_9, \, -z_i+z_j+z_i z_j- z_i z_8- z_i z_9 + z_8 z_9,  \\
&z_i z_j-z_i z_7+z_i z_8-z_j z_8 + z_i z_8 z_9-z_7 z_8 z_9 \} \, \cup \\
&\{-z_3 z_4+z_3 z_7+z_4 z_5-z_4 z_7+z_4 z_8-z_5 z_8+z_3 z_4 z_5-z_3 z_4 z_7-z_3 z_4 z_9-z_3 z_5 z_8+z_3 z_7 z_8\\
&+z_3 z_7 z_9+z_4 z_8 z_9-z_7 z_8 z_9, \, \\
&-z_3 z_5+z_4 z_5+z_3 z_4 z_5-z_3 z_4 z_7+z_3 z_4 z_8-z_3 z_5 z_8-z_3 z_5 z_9-z_3 z_8 z_9+z_4 z_7 z_9+z_5 z_8 z_9 \\
&+z_3 z_4 z_8 z_9-z_3 z_7 z_8 z_9-z_3 z_8 z_9^2+z_7 z_8 z_9^2
\}
\,.
\label{E6alphabet}
}
In Sec. \ref{new}, we shall derive a simpler alphabet by a different gauge choice.

\section{The boundary structure of cluster configuration spaces}
\begin{figure}[htbp]
\center
\includegraphics[scale=0.1]{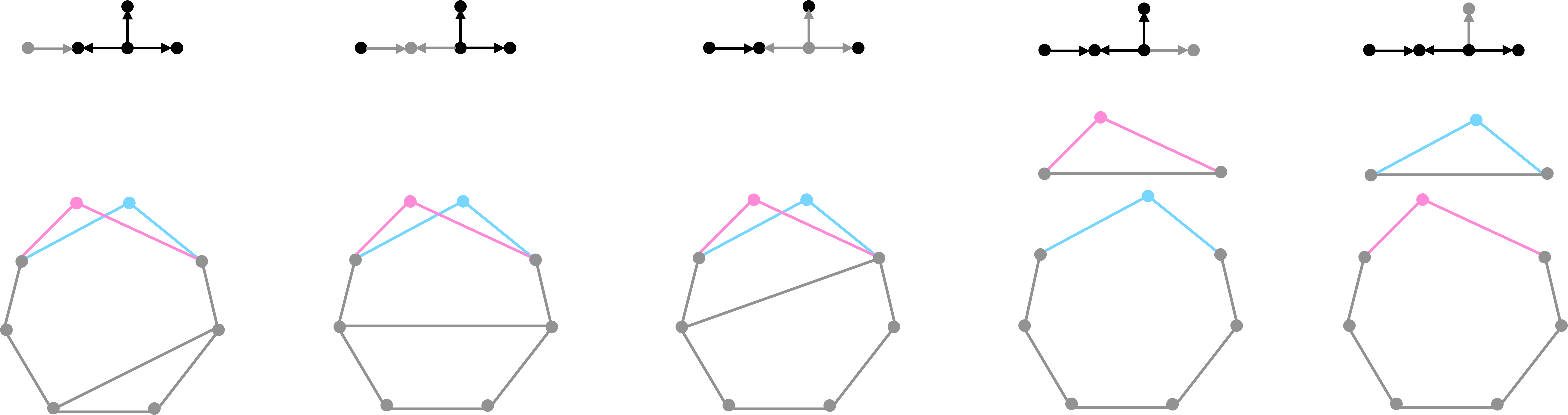}
\caption{The factorizations of the $D_5$ stringy integral as seen on the worldsheet.}
\label{fig:D5boundary}
\end{figure}

Recall that one of the main features of the $A_{n-3}$ worldsheet is that each diagonal divides an $n$-gon into an $(n-k+1)$-gon and a $(k+1)$-gon. The string amplitude factorizes at each pole (boundary of the $u$ space) as
\eq{A_{n-k-2} \times A_{k-2} \subset \partial A_{n-3}  \,.}

We can make similar statements for the other types with the picture of glued polygons.
Each diagonal in the initial triangulation corresponds to a node on the Dynkin diagram and slices the polygon into two parts. The $D_5$ example is shown in Fig. \ref{fig:D5boundary}.
\eq{
\partial D_n=n \left(\sum _{i=1}^{n-2} A_{i-1} \times D_{n-i}+2 A_{n-1}\right) \,.
}
The multiplicity $n$ is determined by the periodicity $n$ of the $D_n$ $Y$-system.

\begin{figure}[htbp]
\center
\includegraphics[scale=0.09]{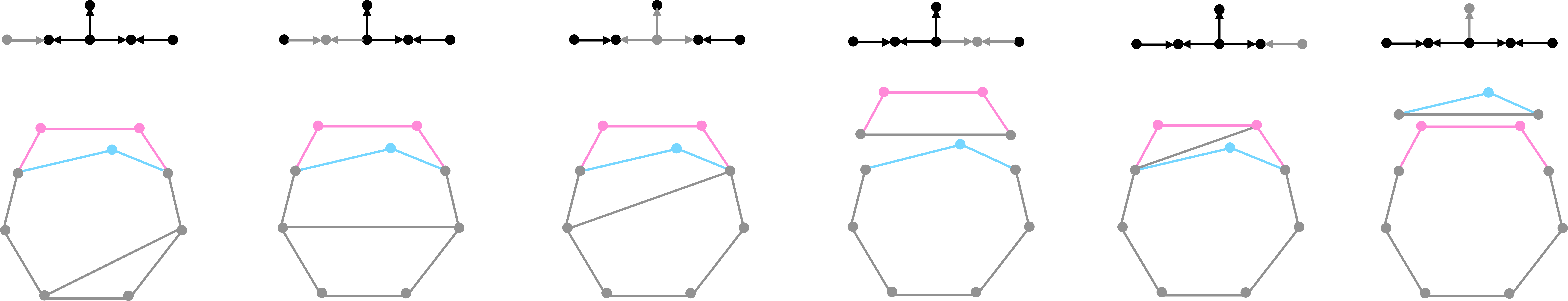}
\caption{The factorizations of the $E_6$ stringy integral as seen on the worldsheet.}
\label{fig:E6boundary}
\end{figure}
The boundaries of the $E_n$ $u$ space can be obtained similarly, as shown in Fig. \ref{fig:E6boundary} for $E_6$.
\eq{
\partial E_6 =  7 \left(A_1 \times A_2 \times A_2+2 A_1 \times A_4+A_5+2 D_5\right) \,.
}
There is an overall factor because the $E_6$ system has period 7. Note that for $E_n$ types, there are more boundaries than diagonals available. We will only identify the possible types of boundaries using the initial cluster, and the remaining boundaries will be obtained by evolving the $Y$-system equations. All the $u_{i}(t)$ at the same $i$th node correspond to the same type of boundary.
For example, the 128 boundaries of the $E_8$ worldsheet are
\eq{
\partial E_8 = 16 \big(A_2 \times D_5+A_1 \times E_6+A_1 \times A_2 \times A_4+A_3 \times A_4+A_1 \times A_6+A_7+D_7+E_7\big) \,.
}

\section{Nonsimply laced types from folding}

\begin{figure}
\center
\includegraphics[scale=0.18]{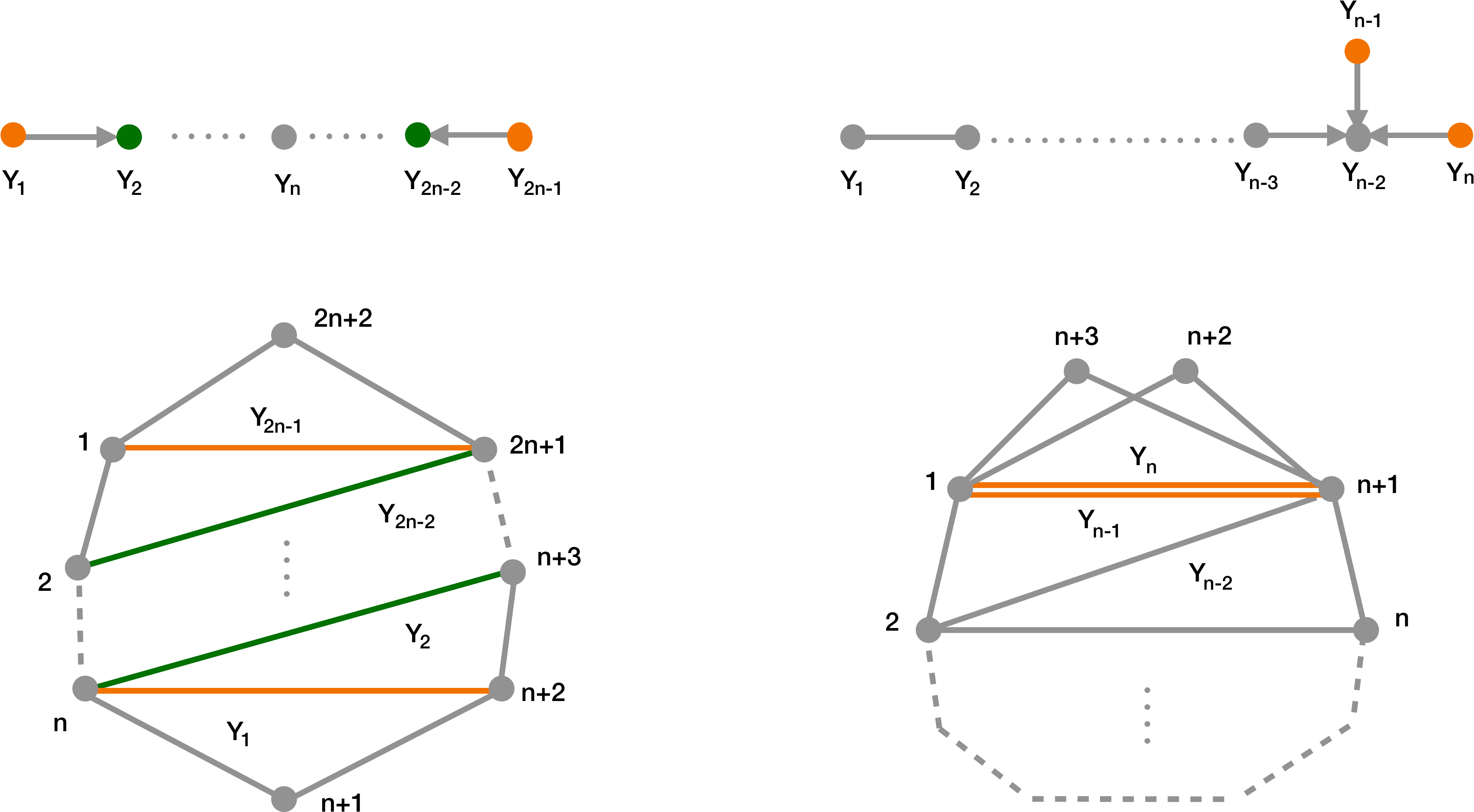}
\caption{The folding from $A_{2n-1}$ to $C_{n}$ and $D_{n}$ to $B_{n-1}$.}
\label{ACDB}
\end{figure}

\begin{figure}
\center
\includegraphics[scale=0.2]{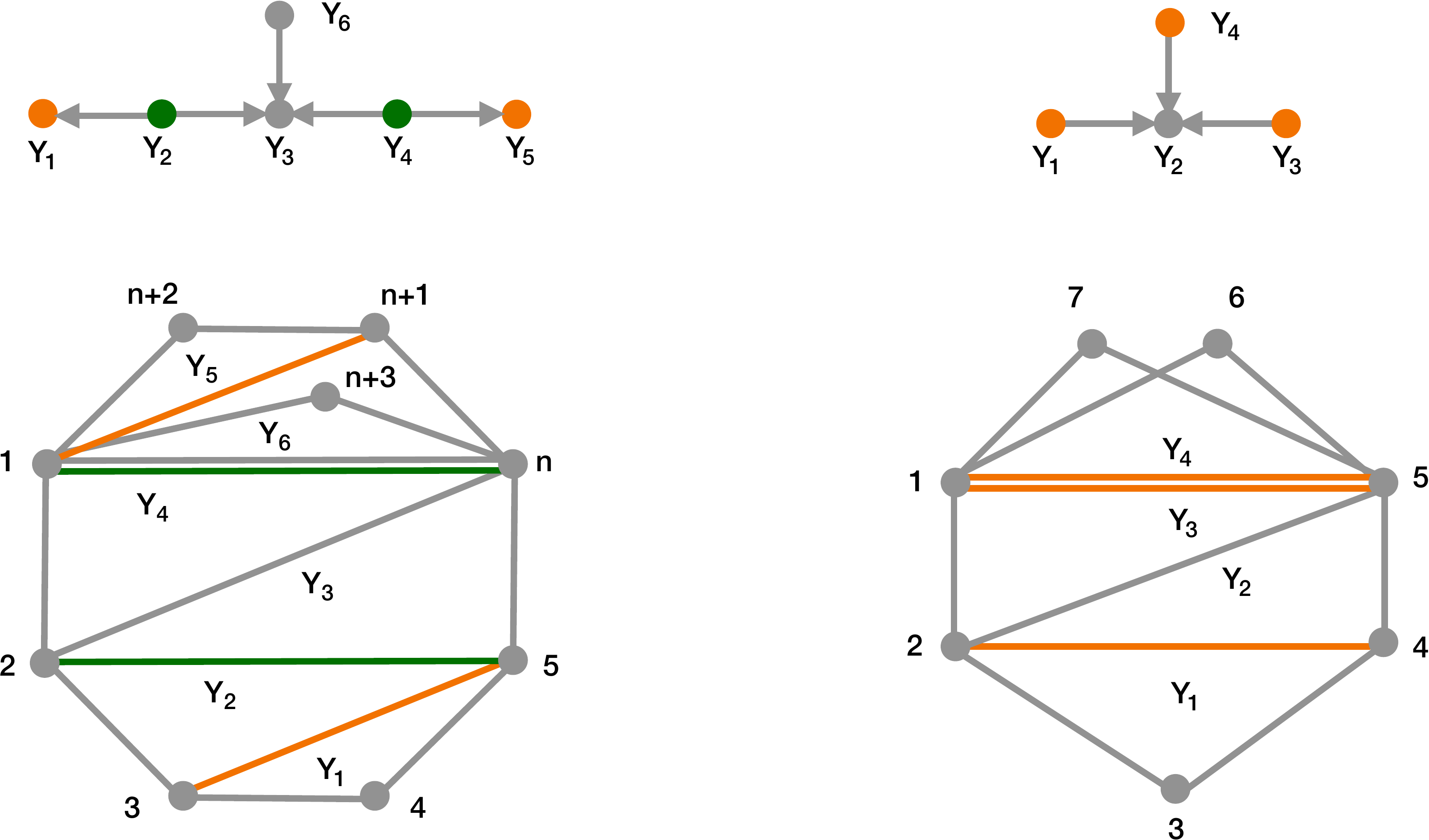}
\caption{The folding from $E_6$ to $F_4$ and $D_4$ to $G_2$.}
\label{EFDG}
\end{figure}

The worldsheet parametrization for the nonsimply laced types can be achieved
by a process known as folding. The folding map on the $z$ parameters is derived from the standard folding of the root systems combined with the birational map in the $ADE$ types. See Figs. \ref{ACDB} and \ref{EFDG}.

\paragraph{$A_{2n-1} \to C_{n}$ folding}
To obtain the $C_{n}$ worldsheet, we fold the $A_{2n-1}$ worldsheet by identifying the diagonals according to the roots as 
\eq{Y_{2n-1} = Y_1, \quad Y_{2n-2} = Y_2, \quad \cdots \quad Y_{n+1} = Y_{n-1}\,.}
Solving the cross-ratio relations, we obtain a fractional map of the $A_{2n-1}$ worldsheet variables in terms of the $C_n$ worldsheet variables:
\eq{
z_{2n+3-i} = \frac{z_{n+3} z_{1,2} z_{n+2-i,n+2}-z_1 z_{2,n+1} z_{n+2,n+3}}{z_{1,2} z_{n+2-i,n+2} - z_{2,n+1} z_{n+2,n+3}} \,. \label{AtoCunfixed}
}
for $i = 1, 2, \cdots, n-1$.
In the standard gauge choice ($z_1 \to -1, z_2 \to 0, z_{n+3} \to \infty$), the folding map \eqref{AtoCunfixed} reduces to a simple gauge-fixed map
\eq{
z_{2n+3-i} = -\frac{z_{n+2}}{z_{n+2-i}}\,. \label{AtoCfixed}
}
We recover the quadratic $C_n$ alphabet from the linear $A_{2n-1}$ alphabet by examining all the polynomial factors that appear in the cross ratio \eqref{crossratio} under the folding map. Equivalently, one may perform the folding map directly on the alphabet \eqref{ACDalphabet} and read off all the factors.

\paragraph{$D_{n} \to B_{n-1}$ folding}
We identify 
\eq{Y_{n} = Y_{n-1} \label{BtoD} \,.}
This is equivalent to $z_{n+3} = z_{n+2}$. 

\paragraph{$E_6 \to F_4$ folding}
To obtain the $F_4$ worldsheet, we fold the $E_6$ worldsheet by identifying the diagonals according to the roots as 
\eq{Y_5 = Y_1, \qquad Y_4 = Y_2\,.}
Solving the cross-ratio relations, we obtain
\eq{
z_7 = \frac{z_6 z_{1,2} z_{3,5}-z_1 z_{2,3} z_{5,6}}{z_{1,2} z_{3,5}-z_{2,3} z_{5,6}},\qquad z_8 = \frac{z_6 z_{1,2} z_{4,5}-z_1 z_{2,4} z_{5,6}}{z_{1,2} z_{4,5}-z_{2,4} z_{5,6}} \,. \label{EtoFunfixed}
}
Upon gauge fixing as $z_1 \to -1, z_2 \to 0, z_6 \to \infty$, this reduces to a simple map
\eq{
z_7 = -\frac{z_5}{z_3}, \qquad z_8 = -\frac{z_5}{z_4} \,. \label{EtoFfixed}
}

\paragraph{$D_4 \to G_2$ folding}
To obtain $G_2$, we fold the $D_4$ worldsheet by identifying the diagonals according to the roots as 
\eq{Y_4 = Y_3 = Y_1 \label{foldG2} \,.}
Solving the cross-ratio relations, we obtain
\eq{
z_6 = \frac{z_5 z_{1,2} z_{3,4}-z_1 z_{2,3} z_{4,5}}{z_{1,2} z_{3,4}-z_{2,3} z_{4,5}} \,. \label{DtoGunfixed}
}
Upon gauge fixing as $z_1 \to -1, z_2 \to 0, z_5 \to \infty$, this reduces to a simple map
\eq{
z_6 = -\frac{z_4}{z_3}\,. \label{DtoGfixed}
}	

The alphabets of nonsimply laced types may be obtained by applying the gauge-fixed maps directly to the alphabets of simply laced types. The results were already quoted in \cite{He:2021zuv} without derivation. In the next section, we shall derive a simpler set of alphabets by first applying the general folding maps and then fixing the gauge.

\section{New cluster alphabets}
\label{new}
In the standard gauge choice, which is usually taken to be $z_1 \to -1, z_2 \to 0, z_{n} \to \infty$, we recover the known cluster alphabets of types $A, C, D$ and obtain new ones for type $E$ and the nonsimply laced ones. However, it is reasonable to suspect that we have not found the simplest possible choice. Unlike the $A_n$ case, not all worldsheet variables are on an equal footing. Different gauge choices will lead to different alphabets. While there is no canonical choice of the alphabet, choosing a gauge that yields letters that are polynomials of the lowest order is preferable. We say two alphabets are equivalent if seen as a collection of hypersurfaces, they have the same topological property. That is,
\begin{enumerate}
\item The number of letters equals the dimension of the cluster algebra.
\item They give the same point count in the hypersurface complement.
\end{enumerate}

A simpler $D_n$ alphabet is obtained by the gauge choice $z_{n+3} \to -1, z_1 \to 0, z_{n+1} \to \infty$. Let 
\eq{a_{i,j} = z_i - z_{n+2} +z_i z_j-z_i z_{n+2}
} 
be the gauge-fixed version of $w_{i,j}$. The $D_n$ alphabet is
\EQ{
\Phi_{D_n}  =  \Phi_{A_{n-1}}(z_2, \ldots, z_{n}) \cup \{z_{n+2}\} \cup \bigcup_{2 \le i \le n} \{z_{i,n+2} \} 
\cup \bigcup_{2 \le i < j \le n} \{a_{i,j}\}\,.}

A similar gauge choice ($z_{9} \to -1, z_1 \to 0, z_{6} \to \infty$) removes the terms containing $z_i^2$ in the $E_6$ alphabet \eqref{E6alphabet}. Let
\EQ{
\widetilde a_{i,j} &= z_i-z_{n+1}+z_i z_j-z_i z_{n+1} \,, \\
a_{i,j,k} &= z_i z_j -z_i z_{n+1}- z_j z_{n+2}+z_{n+1} z_{n+2}+z_i z_j z_k -z_i z_j z_{n+1} -z_i z_k z_{n+2}+z_i z_{n+1} z_{n+2} \,, \\ 
a_{i,j,k,l} &= z_{i} z_{j} -z_{i} z_{n+1}-z_{j} z_{n+2}+z_{n+1} z_{n+2}+z_{i} z_{j} z_{k}+z_{i} z_{j} z_{l} -z_{i} z_{j} z_{n+1} -z_{i} z_{j} z_{n+2} \\ & \quad -z_{i} z_{k} z_{n+1}+z_{i} z_{k} z_{n+2}
  -z_{i} z_{l} z_{n+2}+z_{i} z_{n+1} z_{n+2}-z_{j} z_{k} z_{n+2} +z_{j} z_{n+1} z_{n+2}\\ & \quad+z_{i} z_{j} z_{k} z_{l}  -z_{i} z_{j} z_{k} z_{n+1} -z_{i} z_{j} z_{l} z_{n+2} +z_{i} z_{j} z_{n+1} z_{n+2}
\,.}
be the gauge-fixed versions of $\widetilde{w}_{i,j}^E, w_{i,j,k}, w_{i,j,k,l},$ respectively. A new $E_6$ alphabet may be written succinctly as
\EQ{
\Phi_{E_6} = \Phi_{A_4} (z_2,\ldots,z_5)\cup \{z_7,z_8,z_{7,8},a_{2,3,4,5} \}\cup \bigcup_{2 \le i \le 5} \{ z_{i,7},z_{i,8} \} \\
\cup\bigcup_{2 \le i < j \le 5} \{a_{i,j}, \widetilde a_{i,j}\} \cup  \bigcup_{2 \le i < j < k \le 5} \{a_{i,j,k} \}\,.
}

\paragraph{Gauge fix then fold vs fold then gauge fix}
To obtain the alphabet for nonsimply laced types, we may apply the gauge-fixed folding maps [(\ref{AtoCfixed}, \ref{BtoD}, \ref{EtoFfixed}, \ref{DtoGfixed})] to the alphabets of simply laced types. However, the gauge choice may not be optimal for the nonsimply laced types. Alternatively, we can also first fold the ungauged alphabets using the general maps [(\ref{AtoCunfixed}, \ref{BtoD}, \ref{EtoFunfixed}, \ref{DtoGunfixed})], and then choose the gauge that produces the nicest alphabets for nonsimply laced types. 

If we fold the $D_n$ alphabet to obtain the $B_{n-1}$ alphabet, then some of the letters are still quadratic. If instead the ungauged $D_n$ alphabet are first folded as $z_{n+3} = z_{n+2}$ and then gauge fixed as $z_1 \to -1, z_{n+1} \to 0, z_{n+2} \to \infty$, then we obtain a linear alphabet
\eq{
\Phi_{B_{n-1}} =  \Phi_{A_{n-1}}(z_2, \ldots, z_{n}) \cup \bigcup_{2 \le i < j \le n} \{1-z_{i, j}\}  
\,.}
We chose this particular gauge because it produces the simplest possible set of linear letters. Here we see the advantage of having an ungauged description of the worldsheet. It is not possible to obtain the linear $B_{n-1}$ alphabet from folding the quadratic $D_n$ alphabet because we have gauge fixed $z_{n+1} \to \infty$ to obtain the $D_n$ alphabet, whereas the $B_{n-1}$ alphabet corresponds to gauge fixing $z_{n+2}, z_{n+3} \to \infty$. This shows that the $B$-type alphabet may be realized by a set of hyperplanes.

For $C_n$, we were not able to find a gauge choice that leads to a linear alphabet. The ungauged $C_n$ letters obtained from the general folding map are quadratic in $n-1$ of the variables. An interesting question would be to find a map that realizes the $C_n$ alphabet as a hyperplane arrangement. 

To obtain the $G_2$ alphabet, we now need to solve the folding equation \eqref{foldG2} for $z_2$, apply the map on the ungauged $D_4$ alphabet, and then gauge fix as $z_{1} \to -1, z_5 \to 0, z_{6} \to \infty$. We then arrive at a new $G_2$ alphabet that is at most quadratic, as opposed to quartic as found previously from the standard gauge fixing \cite{He:2021zuv}:
\EQ{
\Phi_{G_2}&=\Phi_{B_2}(z_3, z_4) \cup \left\{z_3-z_4- z_3 z_4 ,\,z_4 - z_3^2+z_3 z_4\right\} 
\,.
}

Note that as in the $D_n \to B_{n-1}$ folding, we cannot obtain a simplified form of the $F_4$ alphabet if we directly folded the $E_6$ alphabet as \eqref{EtoFfixed}, where $z_6$ is already fixed at infinity. Instead, we fold the ungauged alphabet and then gauge fix as $z_{9} \to -1, z_1 \to 0, z_{5} \to \infty$. A new $F_4$ alphabet consists of 28 letters of polynomial degree of at most $4$: 
\EQ{
&\Phi_{F_4} 
= \Phi_{A_4}(z_2, z_3, z_4, z_6) \cup\bigcup_{3\le i \le j \le 4} \left\{a_{i,j}, b_{i,j}, c_{i,j}\right\} \cup \big\{z_2 z_3-z_2 z_4+z_2 z_6 - z_3^2 - z_3^2 z_4 + z_2 z_3 z_6, \\
&\quad\qquad z_2 z_3-z_2 z_4+z_2 z_6-z_3 z_4 -z_3 z_4^2+z_2 z_3 z_6,  z_2^2 z_6 + z_2z_3^2-2 z_2z_3 z_4 +z_2z_3^2 z_6 -z_3^2 z_4^2,\\
&\quad\qquad 2 z_2 z_3-2 z_2 z_4+z_2 z_6 -z_3^2+ z_2 z_3^2+2 z_2 z_3 z_6 -z_2 z_4^2-2  z_3^2z_4+z_2 z_3^2 z_6-z_3^2 z_4^2,\\
&\quad\qquad z_2^2z_6-2 z_2 z_3 z_4 + z_2z_4^2-2 z_2 z_4 z_6+z_2z_6^2  +2 z_3^2 z_4-z_3^2 z_6 +z_2^2 z_6^2-2 z_2 z_3 z_4 z_6+ z_3^2 z_4^2\big\}
\,,}
where
\eq{
a_{i,j} = z_2+z_i z_j, \quad b_{i,j} = z_2 z_6 - z_i z_j, \quad c_{i,j} = z_2+z_6-z_i-z_j +  z_2 z_6 - z_i z_j\,.
}

\paragraph{Acknowledgments}
We thank Song He and Yong Zhang for their collaboration on a related project. Y. W. is supported by China National Natural Science Funds for Distinguished Young Scholar (Grant No. 12105062) and Agence Nationale de la Recherche (ANR), Project No. ANR-22-CE31-0017. P. Z. would like to thank Xiaobin Li and Yuqi Li for the discussions. 

\appendix

\section{Alphabets for $E_7$ and $E_8$}
The alphabets of $E_7$ and $E_8$ may be constructed similarly. The gauge choice is $z_{n+3} \to -1, z_1 \to 0, z_{n} \to \infty$.
The  $E_7$ alphabet consists of letters with a degree of at most $5$.
Introduce two new highest-order letters 
\EQ{
a_{i,j,k,l,m} &= a_{j,m}\, a_{i,k,l}+z_{n+2}\,  z_{i,j}\,  z_{k,n+1}\,  z_{l,m} \,,\\
\widetilde a_{i,j,k,l,m} &= \widetilde{a}_{j,l}\, a_{i,k,m}-z_{n+2} \,  z_{i,j}\,  z_{k,l} \, z_{n+1,n+2} \,.
}
The 70 letters of the $E_7$ alphabet are
\EQ{
\Phi_{E_7} =\Phi_{A_5} (z_2,\ldots,z_6)\cup \{z_8,z_9,z_{8,9},a_{2,3,4,5,6},\widetilde a_{2,3,4,5,6} \} \cup \bigcup_{2 \le i \le 6} \{ z_{i,8},z_{i,9} \} \\
 \cup\bigcup_{2 \le i < j \le 6} \{a_{i,j}, \widetilde a_{i,j}\} \cup  \bigcup_{2 \le i < j < k \le 6} \{a_{i,j,k} \} \cup  \bigcup_{2 \le i < j < k < l \le 6} \{a_{i,j,k,l}\} \,.
}

The $E_8$ alphabet consists of letters with a degree of at most $7$. We introduce six degree-7 letters
\EQ{
c_2
&= a_{2,4,6} a_{2,3,5,7}+z_{2} z_{10} z_{2,3} z_{4,5} z_{6,7} z_{9,10} \,,\\
c_3 
&= a_{3,4,6} a_{2,3,5,7}+z_{10} z_{2,3} z_{3,4} z_{5,6} z_{9,10}  \,, \\
c_4 
&= a_{2,4,7} a_{3,4,5,6}-z_{10}  z_{2,3} z_{4,5} z_{9,10} a_{4,6} \,, \\
c_5
&= a_{2,5,6} a_{3,4,5,7} + z_{2,3} z_{5,9} z_{6,7} z_{10} \widetilde{a}_{4,5}\,, \\
c_6 
&= a_{3,5,6} a_{2,4,6,7}+z_{2} z_{10} z_{3,4} z_{5,6} z_{6,7} z_{9,10} \,, \\
c_7 
&= a_{3,5,7} a_{2,4,6,7}+z_{10} z_{2,3} z_{4,5} z_{6,7} z_{9,10} \,,
}
and three degree-6 letters
\EQ{
b_1 &= a_{3,7} a_{2,4,5,6} + z_{10}  z_{2,3} z_{6,7} a_{4,5} \,, \\
b_2 &= \widetilde{a}_{2,6} a_{3,4,5,7} + z_{10} z_{2,3} z_{6,7} \widetilde{a}_{4,5} \,,\\
b_3 &= a_{2,3,5,6,7} z_{4,10} - z_{3,4} z_{9,10}a_{2,5,6} \,.
}
The 128 letters of the $E_8$ alphabet are
\EQ{
\Phi_{E_8}
=\Phi_{A_6} (z_2,\ldots,z_7)\cup \{z_9,z_{10},z_{9,10},b_1, b_2, b_3 \}\cup \bigcup_{2 \le i \le 7} \{ z_{i,9},z_{i,10}, c_i \}  \cup\bigcup_{2 \le i < j \le 7} \{a_{i,j}, \widetilde a_{i,j}\} \\ 
\cup  \bigcup_{2 \le i < j < k \le 7} \{a_{i,j,k} \}  \cup  \bigcup_{2 \le i < j < k < l \le 7} \{a_{i,j,k,l}\} \cup \bigcup_{2 \le i < j < k < l < m \le 7} \{a_{i,j,k,l,m},\widetilde a_{i,j,k,l,m}\} \,.
}

\bibliographystyle{elsarticle-num} 
\bibliography{bib}

\begin{thebibliography}{10}
\expandafter\ifx\csname url\endcsname\relax
  \def\url#1{\texttt{#1}}\fi
\expandafter\ifx\csname urlprefix\endcsname\relax\def\urlprefix{URL }\fi
\expandafter\ifx\csname href\endcsname\relax
  \def\href#1#2{#2} \def\path#1{#1}\fi

\bibitem{Chan:1969xg}
H.-M. Chan, {A generalised Veneziano model for the $n$-point function}, Phys.
  Lett. B 28 (1969) 425--428.
\newblock \href {https://doi.org/10.1016/0370-2693(69)90342-6}
  {\path{doi:10.1016/0370-2693(69)90342-6}}.

\bibitem{Chan:1969ex}
H.-M. Chan, S.~T. Tsou, {Explicit construction of the $n$-point function in the
  generalized Veneziano model}, Phys. Lett. B 28 (1969) 485--488.
\newblock \href {https://doi.org/10.1016/0370-2693(69)90523-1}
  {\path{doi:10.1016/0370-2693(69)90523-1}}.

\bibitem{Koba:1969rw}
Z.~Koba, H.~B. Nielsen, {Reaction amplitude for $n$ mesons: A Generalization of
  the Veneziano-Bardakci-Ruegg-Virasoro model}, Nucl. Phys. B 10 (1969)
  633--655.
\newblock \href {https://doi.org/10.1016/0550-3213(69)90331-9}
  {\path{doi:10.1016/0550-3213(69)90331-9}}.

\bibitem{Koba:1969kh}
Z.~Koba, H.~B. Nielsen, {Manifestly crossing invariant parametrization of $n$
  meson amplitude}, Nucl. Phys. B 12 (1969) 517--536.
\newblock \href {https://doi.org/10.1016/0550-3213(69)90071-6}
  {\path{doi:10.1016/0550-3213(69)90071-6}}.

\bibitem{Arkani-Hamed:2019mrd}
N.~Arkani-Hamed, S.~He, T.~Lam, {Stringy canonical forms}, JHEP 02 (2021) 069.
\newblock \href {http://arxiv.org/abs/1912.08707} {\path{arXiv:1912.08707}},
  \href {https://doi.org/10.1007/JHEP02(2021)069}
  {\path{doi:10.1007/JHEP02(2021)069}}.

\bibitem{Arkani-Hamed:2019plo}
N.~Arkani-Hamed, S.~He, T.~Lam, H.~Thomas, {Binary geometries, generalized
  particles and strings, and cluster algebras}, Phys. Rev. D 107~(6) (2023)
  066015.
\newblock \href {http://arxiv.org/abs/1912.11764} {\path{arXiv:1912.11764}},
  \href {https://doi.org/10.1103/PhysRevD.107.066015}
  {\path{doi:10.1103/PhysRevD.107.066015}}.

\bibitem{zbMATH07431231}
N.~{Arkani-Hamed}, S.~{He}, T.~{Lam}, {Cluster configuration spaces of finite
  type}, {SIGMA, Symmetry Integrability Geom. Methods Appl.} 17 (2021) paper
  092, 41.
\newblock \href {https://doi.org/10.3842/SIGMA.2021.092}
  {\path{doi:10.3842/SIGMA.2021.092}}.

\bibitem{zbMATH02068688}
S.~{Fomin}, A.~{Zelevinsky}, {\(Y\)-systems and generalized associahedra},
  {Ann. Math. (2)} 158~(3) (2003) 977--1018.
\newblock \href {https://doi.org/10.4007/annals.2003.158.977}
  {\path{doi:10.4007/annals.2003.158.977}}.

\bibitem{Dixon:2011pw}
L.~J. Dixon, J.~M. Drummond, J.~M. Henn, {Bootstrapping the three-loop
  hexagon}, JHEP 11 (2011) 023.
\newblock \href {http://arxiv.org/abs/1108.4461} {\path{arXiv:1108.4461}},
  \href {https://doi.org/10.1007/JHEP11(2011)023}
  {\path{doi:10.1007/JHEP11(2011)023}}.

\bibitem{Golden:2013xva}
J.~Golden, A.~B. Goncharov, M.~Spradlin, C.~Vergu, A.~Volovich, {Motivic
  Amplitudes and Cluster Coordinates}, JHEP 01 (2014) 091.
\newblock \href {http://arxiv.org/abs/1305.1617} {\path{arXiv:1305.1617}},
  \href {https://doi.org/10.1007/JHEP01(2014)091}
  {\path{doi:10.1007/JHEP01(2014)091}}.

\bibitem{Drummond:2014ffa}
J.~M. Drummond, G.~Papathanasiou, M.~Spradlin, {A Symbol of Uniqueness: The
  Cluster Bootstrap for the 3-Loop MHV Heptagon}, JHEP 03 (2015) 072.
\newblock \href {http://arxiv.org/abs/1412.3763} {\path{arXiv:1412.3763}},
  \href {https://doi.org/10.1007/JHEP03(2015)072}
  {\path{doi:10.1007/JHEP03(2015)072}}.

\bibitem{Chicherin:2020umh}
D.~Chicherin, J.~M. Henn, G.~Papathanasiou, {Cluster algebras for Feynman
  integrals}, Phys. Rev. Lett. 126~(9) (2021) 091603.
\newblock \href {http://arxiv.org/abs/2012.12285} {\path{arXiv:2012.12285}},
  \href {https://doi.org/10.1103/PhysRevLett.126.091603}
  {\path{doi:10.1103/PhysRevLett.126.091603}}.

\bibitem{He:2021zuv}
S.~He, Y.~Wang, Y.~Zhang, P.~Zhao, {Notes on Worldsheet-Like Variables for
  Cluster Configuration Spaces}, SIGMA 19 (2023) 045.
\newblock \href {http://arxiv.org/abs/2109.13900} {\path{arXiv:2109.13900}},
  \href {https://doi.org/10.3842/SIGMA.2023.045}
  {\path{doi:10.3842/SIGMA.2023.045}}.

\bibitem{Zamolodchikov:1991et}
A.~B. Zamolodchikov, {On the thermodynamic Bethe ansatz equations for
  reflectionless ADE scattering theories}, Phys. Lett. B 253 (1991) 391--394.
\newblock \href {https://doi.org/10.1016/0370-2693(91)91737-G}
  {\path{doi:10.1016/0370-2693(91)91737-G}}.

\bibitem{Gekhtman_2005}
M.~Gekhtman, M.~Shapiro, A.~Vainshtein,
  \href{https://doi.org/10.1215%2Fs0012-7094-04-12723-x}{{Cluster algebras and
  Weil-Petersson forms}}, Duke Mathematical Journal 127~(2) (apr 2005).
\newblock \href {https://doi.org/10.1215/s0012-7094-04-12723-x}
  {\path{doi:10.1215/s0012-7094-04-12723-x}}.
\newline\urlprefix\url{https://doi.org/10.1215%2Fs0012-7094-04-12723-x}

\bibitem{Volkov_2007}
A.~Y. Volkov, \href{https://doi.org/10.1007%2Fs00220-007-0343-y}{{On the
  Periodicity Conjecture for $Y$-systems}}, Communications in Mathematical
  Physics 276~(2) (2007) 509--517.
\newblock \href {https://doi.org/10.1007/s00220-007-0343-y}
  {\path{doi:10.1007/s00220-007-0343-y}}.
\newline\urlprefix\url{https://doi.org/10.1007%2Fs00220-007-0343-y}

\bibitem{Drummond:2018dfd}
J.~Drummond, J.~Foster, O.~G\"urdo\u{g}an, {Cluster adjacency beyond MHV}, JHEP
  03 (2019) 086.
\newblock \href {http://arxiv.org/abs/1810.08149} {\path{arXiv:1810.08149}},
  \href {https://doi.org/10.1007/JHEP03(2019)086}
  {\path{doi:10.1007/JHEP03(2019)086}}.

\end{thebibliography}

\end{document}